\documentclass[usenatbib]{mn2e} \usepackage{psfig}\usepackage{graphicx}\usepackage{amsmath}\usepackage{amssymb}\usepackage{rotating}\usepackage{lscape} \usepackage{psfig}\usepackage{graphicx}\usepackage{amsmath}\usepackage{amssymb}\usepackage{rotating}\usepackage{lscape}

\title[{\it XMM-Newton} discovery of transient X-ray pulsar in M33]{XMM-Newton discovery of transient 285.4 s X-ray pulsar XMMU J013359.5+303634 in M33}

\author[S.\,P.\,Trudolyubov] {S.\,P.\,Trudolyubov\thanks{E-mail: strudolyubov@gmail.com}\\ Predictive Analytics, 2935 Rodeo Park Drive E., Santa Fe, NM 87505, USA\\}

\date{}

\pagerange{\pageref{firstpage}--\pageref{lastpage}} \pubyear{2013}

\begin{document}

\maketitle

\label{firstpage}

\begin{abstract}
I report on the discovery and analysis of the first transient X-ray pulsar detected in the Local Group galaxy M33. The 2010 
July-August deep {\em XMM-Newton} observations of M33 fields revealed a new bright X-ray source XMMU J013359.5+303634 exhibiting pulsations 
with a period $P \sim 285.4$ s and pulsed fraction $\sim$47 per cent in the 0.3-10 keV energy range. The pulse phase averaged spectrum of 
XMMU J013359.5+303634 is typical of X-ray pulsars and can be fit with an absorbed simple power law model of photon index $\Gamma \sim 1.2$ 
in the 0.3-10 keV energy band. The search for an optical counterpart did not yield any stellar object brighter than 20 mag, suggesting that 
XMMU J013359.5+303634 is not a Galactic foreground object and almost certainly belongs to M33. Assuming the distance of 817 kpc, the maximum 
observed luminosity of the source in the 0.3-10 keV energy range is $\sim 1.4\times10^{37}$ ergs s$^{-1}$, at least 20 times higher than 
quiescent luminosity. The brightest optical object inside the error circle of XMMU J013359.5+303634 has a visual magnitude of 20.9 and 
properties consistent with being an early B V star when placed at a distance of M33. Based on the X-ray pulsations and spectrum, transient 
behavior and possible early B class optical counterpart, XMMU J013359.5+303634 can be classified as another extragalactic Be/X-ray binary 
candidate. 
\end{abstract}


\section{Introduction}
Since their discovery \citep{Giacconi71}, accreting X-ray pulsars have been prime targets for both observational and theoretical study 
\citep{WSH83,Nagase89,Bildsten97}. X-ray pulsars proved to be unique laboratories that allow to test a wide variety of physical processes 
in the presence of magnetic fields and strong gravity, and study stellar evolution in binary systems. 

Until recently, the study of pulsating X-ray sources was limited to our Galaxy and the nearby Magellanic Clouds. The advanced capabilities 
of {\em Chandra} and {\em XMM-Newton} observatories opened the possibility to study spectral and timing properties of individual bright 
X-ray sources associated with more distant galaxies \citep{F06,FW06}, allowing to extend searches for pulsating X-ray binaries beyond the 
Milky Way and its immediate neighbors. Recent analysis of the monitoring observations with {\em XMM-Newton} and {\em Chandra} has already 
led to the discovery of coherent pulsations in three X-ray sources located in the Local Group galaxy M31 \citep{O01,T05,TP08} and two X-ray 
pulsars in the nearby spiral galaxies NGC 2403 and NGC 1313 \citep{TPC07,T08}. Two of the pulsating sources detected in M31 have supersoft 
spectra that can be fit with blackbody models with characteristic temperatures of several tens of eV, and probably belong to an accreting 
white dwarf systems. The remaining three objects have hard spectra with photon index $\Gamma \sim 0.9 - 1.5$, typical of systems 
containing accreting highly-magnetized neutron star. These results clearly demonstrate that bright ($L_{X} \gtrsim 10^{37}$ ergs s$^{-1}$) 
pulsating X-ray source  with both supersoft and hard spectra can be detected in the galaxies inside and beyond the Local Group up to the 
distances of a few Mpc.  

At the distance of 817 kpc \citep{Freedman01}, the nearest late-type face-on spiral galaxy M33 provides a good opportunity to 
study X-ray sources associated with a young stellar population. M33 was the target of multiple X-ray observations with the {\em Einstein} 
\citep{Trinchieri88}, {\em ROSAT} \citep{HP01}, {\em Chandra} \citep{Plucinsky08,Tullmann11}, and {\em XMM-Newton} observatories 
\citep{Pietsch04,Misanovic06}. These observations uncovered a population of hundreds of X-ray sources with many of them identified as 
X-ray binary and supernova remnant candidates associated with regions of recent star formation. 

In this paper, I report on the discovery of the coherent 285.4 s pulsations in the flux of transient X-ray source XMMU J013359.5+303634 
in M33, using archival data of {\em XMM-Newton} observations. We study X-ray properties of the source, search for the optical counterpart 
and discuss its nature.

\begin{table*}
\caption{{\em XMM-Newton} observations of M33 used in the analysis of XMMU J013359.5+303634. 
\label{obslog}}
\begin{tabular}{cccccccl}
\hline
Date     & Obs. ID  & Instrument & Mode/  & RA (J2000)$^{a}$ & Dec (J2000)$^{a}$ & Exp.$^{b}$\\
         &          &            & Filter & (h:m:s)          & (d:m:s)           & (ks)      \\
\hline
2010 Jul. 9-10  & 0650510101 & EPIC  & Full/Medium  & 01:34:13.40 & +30:47:48.8 & 99.7\\
2010 Jul. 11-12 & 0650510201 & EPIC  & Full/Medium  & 01:33:45.22 & +30:35:44.4 & 99.6\\
2010 Aug. 12-13 & 0650510601 & EPIC  & Full/Medium  & 01:34:21.29 & +30:27:07.3 & 100.0\\
\hline  
\end{tabular}

$^{a}$ -- pointing coordinates\\
$^{b}$ -- instrument exposure used in the analysis\\

\end{table*}

\section{Observations and data reduction} 
In the following analysis I used the data of the 2010 July 9-10, July 11-12 and August 12-13 {\em XMM-Newton} observations of M33 fields obtained 
by the  three European Photon Imaging Camera (EPIC) instruments (MOS1, MOS2 and pn) \citep{Turner01,Strueder01} (Table \ref{obslog}). Several 
{\em Chandra} (Obs. IDs: 786, 1730, 6376, 6377, 6385, 6387, 6388, 6389, 7170, 7171, 7196) \citep{Plucinsky08,Tullmann11} observations 
with the field of view covering the position of XMMU J013359.5+303634 were also used to obtain upper limits on the source quiescent flux.

The {\em XMM-Newton}/EPIC data files were processed using the {\em XMM-Newton} Science Analysis System (SAS v 11.0.0)\footnote{See http://xmm.esa.int/sas}. 
The original event files were screened to exclude time intervals with high background levels. For timing analysis, the EPIC event arrival 
times were corrected to the solar system barycenter using the SAS task {\em barycen}.  

Point sources in the EPIC-pn and MOS images of the M33 fields were detected and localized using the SAS maximum likelihood source detection 
script {\em edetect\_chain}. Bright X-ray sources with known counterparts from USNO-B and LGGS catalogs \citep{Monet03,Massey06} were 
used to correct the EPIC image astrometry. The {\em XMM-Newton} source positions were also cross-correlated with {\em Chandra} source lists. 
The residual systematic error in the corrected source positions is estimated to be of the order $\sim 1$ arcsec.

To generate energy spectra and light curves of XMMU J013359.5+303634, the source counts were extracted from elliptical or circular regions 
including at least $\sim 70$ per cent of the source energy flux. For the 2010 Jul. 9-10 observation, the elliptical extraction regions with 
semi-axes of 20$\arcsec$ and 12$\arcsec$ and position angle of 50$^{\circ}$ (EPIC-pn) and 20$^{\circ}$ (EPIC-MOS) were used. For the 2010 Jul. 11-12, 
circular regions of 18$\arcsec$ radius were used to extract source spectra and light curves from EPIC-pn and MOS detectors. Adjacent 
source-free regions of the same size were used to extract background spectra and light curves. The source and background spectra were then 
renormalized applying the standard SAS tasks. The spectra and light curves were built by selecting valid single and double (pattern 0-4) events 
for EPIC-pn and single-quadruple (pattern 0-12) events for the MOS cameras in the $0.3 - 10$ keV energy range. To improve timing analysis sensitivity, 
the combined source and background light curves from individual EPIC detectors were used. The EPIC-pn and MOS light curves were extracted using 
identical time filtering criteria based on  Mission Relative Time (MRT), following the procedure described in \cite{Robin_timing} to ensure 
proper synchronization. 

For the 2010 July 9-10 and July 11-12 observations, EPIC-pn and MOS source spectra were grouped to contain a minimum of 20 counts per spectral 
bin, in order to allow use of $\chi^{2}$ statistics, and fit to analytic models using the XSPEC 
v.12\footnote{http://heasarc.gsfc.nasa.gov/docs/xanadu/xspec/index.html} fitting package \citep{arnaud96}. For the 2010 Aug. 12-13 observation, 
the source count rate was not sufficient to perform spectral analysis. The spectra from individual EPIC detectors were fitted simultaneously 
with independently varying normalizations. For the timing analysis, the standard XANADU/XRONOS 
v.5\footnote{http://heasarc.gsfc.nasa.gov/docs/xanadu/xronos/xronos.html} tasks and Lomb-Scargle period search method were used 
\citep{Scargle82,Press07}.

The data of the {\em Chandra} observations were processed using the CIAO v4.4\footnote{http://asc.harvard.edu/ciao/} threads. The standard 
screening of the {\em Chandra} data was used to exclude time intervals with high background levels. For each observation, the X-ray images 
in the 0.3-7 keV energy band were generated, and the CIAO wavelet detection routine {\em wavdetect} was used to detect point sources.

To estimate upper limits on the quiescent source luminosity, the 2$\sigma$ upper limits on {\em Chandra}/ACIS and {\em XMM-Newton}/EPIC 
count rates estimated from the X-ray images, were converted into energy fluxes in the 0.3-10 keV energy range using Web 
PIMMS\footnote{http://heasarc.gsfc.nasa.gov/Tools/w3pimms.html}, assuming an absorbed power law spectral shape with photon index $\Gamma = 1.2$ 
and foreground absorbing column $N_{\rm H}$=$1.5\times10^{20}$ cm$^{-2}$. For the 2010 Aug. 12-13 observation, the same procedure was applied to 
convert source count rate from {\em XMM-Newton}/EPIC images to luminosity.  

In the following analysis, a distance of 817 kpc is assumed for M33 \citep{Freedman01}. All parameter errors quoted are 68\% ($1\sigma$) 
confidence limits.

\section{Results}
\subsection{Source detection and possible optical counterparts}
\begin{figure}
\begin{tabular}{c}
\psfig{figure=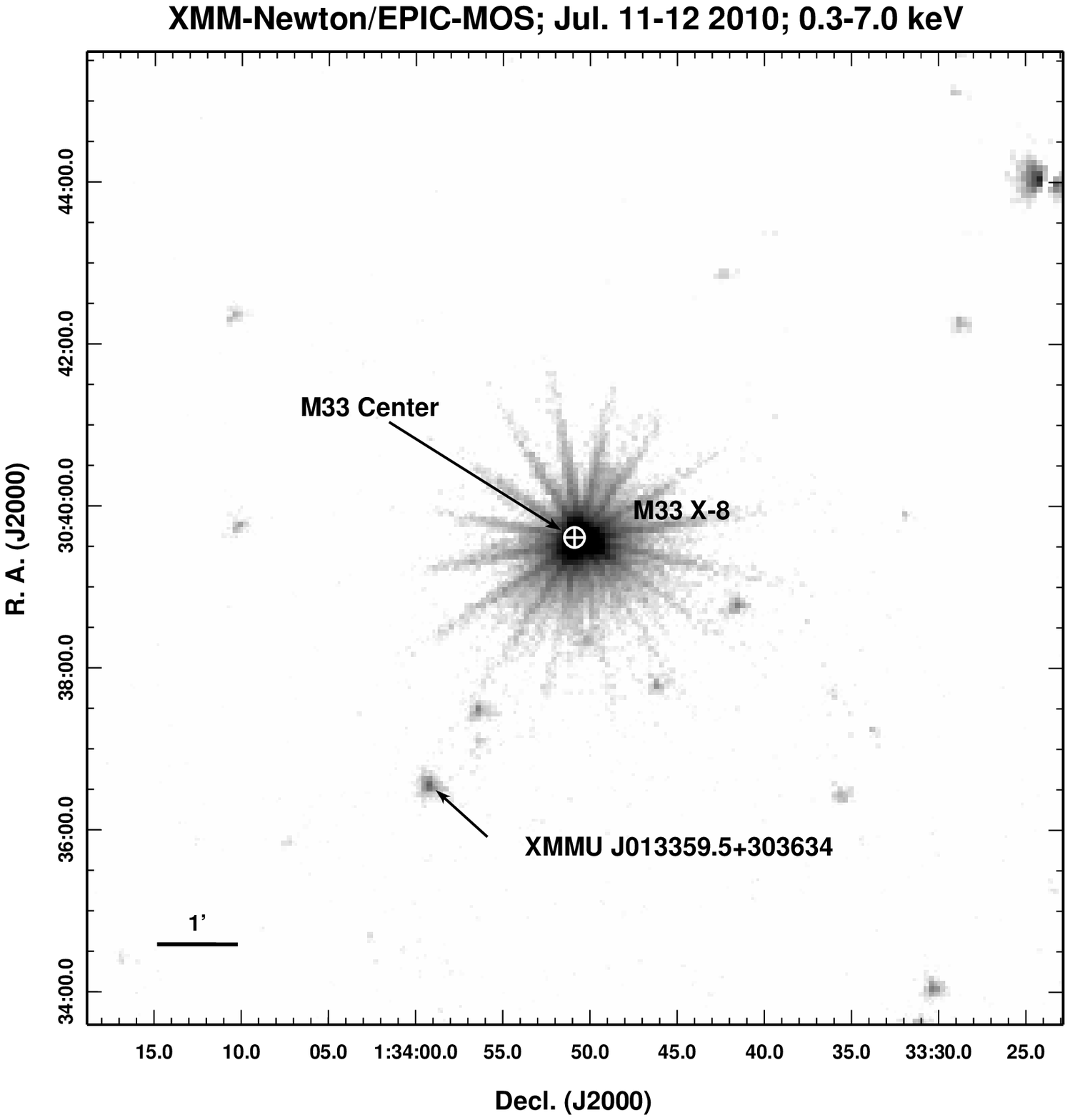,width=8.0cm,angle=0.}\\
\\
\psfig{figure=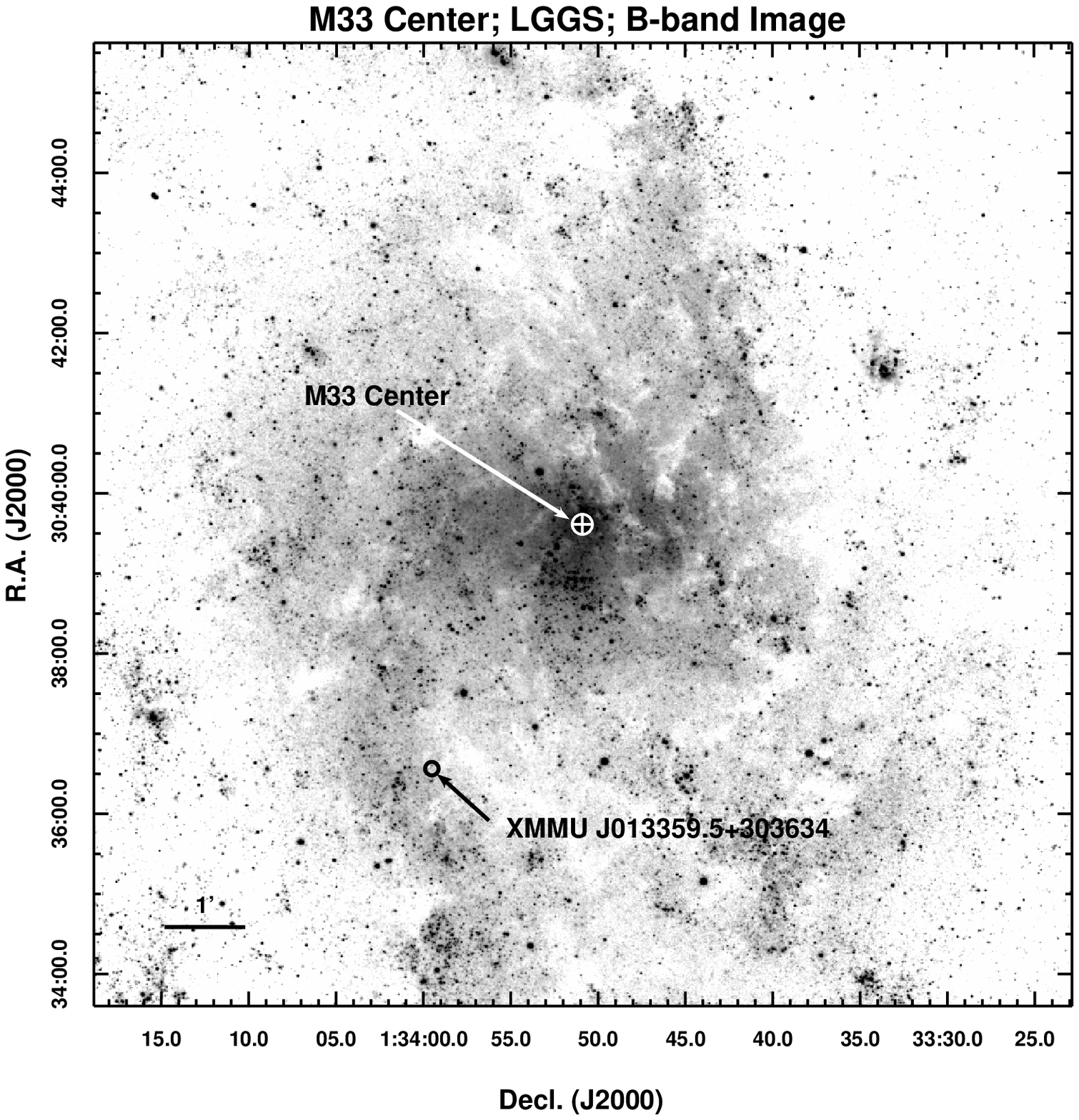,width=8.5cm,angle=0.}
\end{tabular}
\caption{Upper panel: Combined 0.3-7 keV {\em XMM}/EPIC-MOS image covering a 12$\arcmin\times12\arcmin$ central region of M33, 
taken on Jun. 11-12, 2010. The position of the new pulsar XMMU J013359.5-303634 is marked with an arrow, and the position of 
M33 nucleus is shown with a white cross. Lower panel: Optical (B-band) image of the region covered by the X-ray image from the 
Local Group Galaxies Survey \citep{Massey06}. The position of XMMU J013359.5-303634 is shown with black circle of $5\arcsec$ 
radius.}\label{image_general}
\end{figure}

\begin{figure}
\psfig{figure=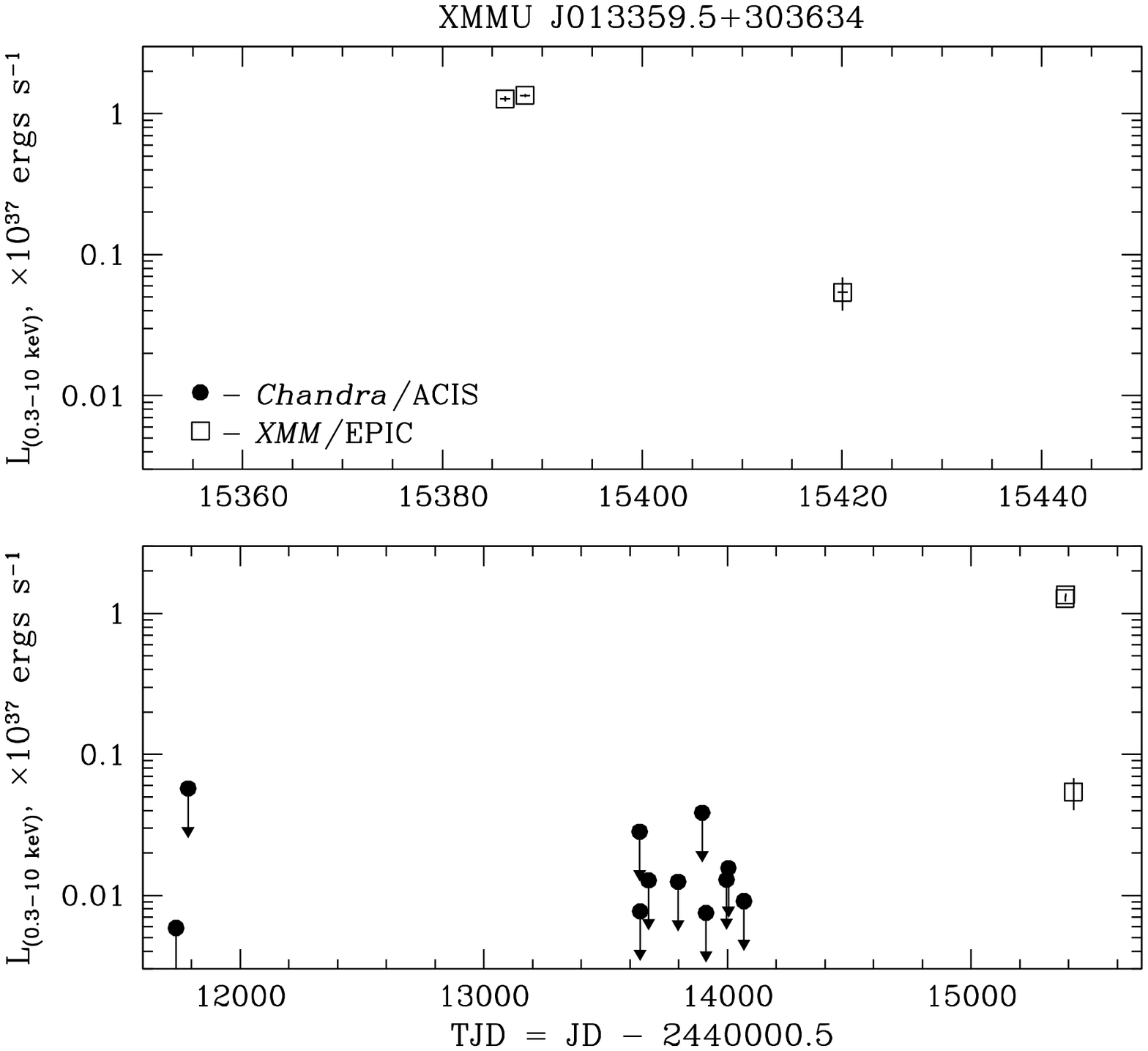,width=9.0cm,angle=0.}
\caption{Upper panel: X-ray light curve of XMMU J013359.5+303634 in the 0.3-10 keV energy band based on the results of 2010 {\em XMM-Newton} 
observations. Lower panel: Long-term light curve of XMMU J013359.5+303634 based on the results of 2000-2010 {\em Chandra} and {\em XMM-Newton} 
observations. The source luminosities were calculated assuming the distance of 817 kpc.}\label{long_term_lc}
\end{figure}

A new transient X-ray source XMMU J013359.5+303634 has been discovered in the data of the 2010 July 9-10 and  11-12 {\em XMM-Newton} observations 
of the M33 field (Table \ref{obslog}). The observed source luminosity estimated from spectral analysis was $\sim 1.4 \times 10^{37}$ ergs s$^{-1}$ 
(Table \ref{timing_spec_par}). XMMU J013359.5+303634 was marginally detected during the subsequent, 2010 August 
12-13 {\em XMM-Newton} observation with a count rate of $(1.3\pm0.3)\times10^{-3}$ cnts s$^{-1}$ in EPIC-pn corresponding to a much lower luminosity 
of $\sim 5.4\times 10^{35}$ ergs s$^{-1}$. Using the corrected EPIC image astrometry, the estimated position of XMMU J013359.5+303634 is 
$\alpha = 01^{h} 33^{m} 59.52^{s}, \delta = +30^{\circ} 36\arcmin 34.5\arcsec$ (J2000 equinox) with an uncertainty of $\sim 1.0\arcsec$ 
(Fig. \ref{image_general}). The projected galactocentric distance of XMMU J013359.5+303634 is $\sim 3.6\arcmin$ or $\sim 850$ pc. 
The analysis of other archival observations of the same field with {\em Chandra} did not yield any source detection with an upper limit (2$\sigma$) 
ranging from $\sim 5.9\times 10^{34}$ to $\sim 5.7\times 10^{35}$ ergs s$^{-1}$ (or $\sim$ 24-230 times lower than the observed outburst luminosity), 
depending on the duration of the observation and instrument used (Fig. \ref{long_term_lc}). In addition, the source was not detected in the 
combined 2001-2003 {\em XMM-Newton} survey observations of M33 \citep{Pietsch04} down to a luminosity level of $\sim 10^{35}$ ergs s$^{-1}$.

\begin{table*}
\caption{Objects from Local Group Galaxies Survey (LGGS) catalog \citep{Massey06} located within the 3$\sigma$ error circle of 
XMMU J013359.5+303634. 
\label{LGS_counterparts}}
\begin{tabular}{ccccccc}
\hline
LGGS ID  & RA (J2000)  & Dec (J2000) & $V$ mag & $(B-V)$& $(U-B)$ & $\Delta$\\
         & (h:m:s)     & (d:m:s)     &         &  (mag) &  (mag)  & ($\arcsec$)\\
\hline
 105063  & 01:33:59.47 & +30:36:33.0 & 22.40 & 0.30 & -0.01 & 1.6 \\
 105149  & 01:33:59.52 & +30:36:36.0 & 20.92 &-0.11 & -0.86 & 1.5 \\
 105311  & 01:33:59.60 & +30:36:34.3 & 21.17 & 0.39 & -0.96 & 1.1 \\
\hline  
\end{tabular}
\end{table*}  

\begin{figure}
\psfig{figure=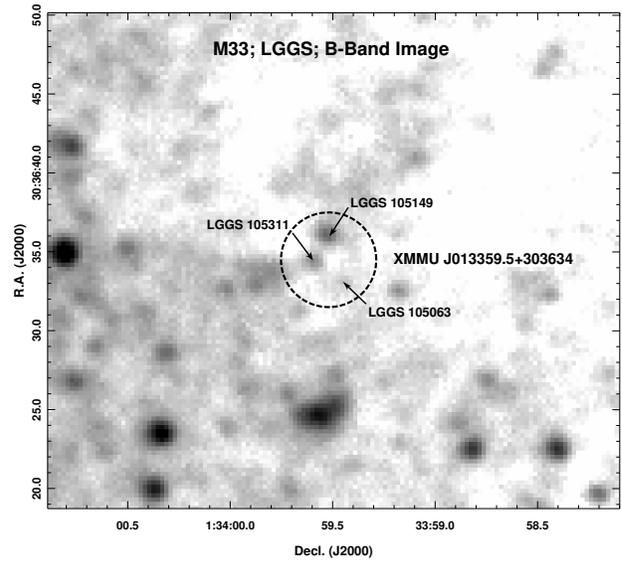,width=8.5cm,angle=0.}
\caption{Optical (B-band) image from the Local Group Galaxies Survey \citep{Massey06} showing three optical objects within 3$\sigma$ error 
circle of XMMU J013359.5-303634 (dashed line).}\label{finding_chart}
\end{figure}

The search for the optical counterpart using the images from the Digitized Sky Survey (red and blue) did not reveal any stellar-like object 
brighter than $\sim 20$ mag within the 3$\sigma$ error circle of XMMU J013359.5+303634. A cross-correlation with optical catalog based on 
the deep, high resolution Local Group Galaxies Survey (LGGS) \citep{Massey06} yielded three possible counterparts of $\sim$21-22 mag in $V$ 
band within 3$\arcsec$ distance from the X-ray position (Fig. \ref{finding_chart}, Table \ref{LGS_counterparts}). Assuming the M33 optical 
color excess $E(B-V)$=0.12 \citep{Massey07}, the brightest optical object (LGGS 105149) has intrinsic colors $(B-V)_{0}\approx-0.2$ and 
$(U-B)_{0}\approx-0.9$ and absolute magnitude $M_{V0} \approx -4$ consistent with being an early B V main sequence star 
\citep{Massey06,Cox00}. It also falls into the Be star domain in the $(U-B)/(B-V)$ diagram, as does its reddening-free Johnson Q-index of 
$\sim -0.8$ \citep{FM79,Massey07}. The second brightest optical object (LGGS 105311) with estimated absolute magnitude of $\sim -3.5$ is 
probably a blend of two or more unresolved stars of different spectral class, having $(U-B)_{0}\sim-1.0$ typical of O to B star and 
$(B-V)_{0}\sim0.3$ of late A to early F spectral class \citep{Cox00}. The third object (LGGS 105063) shows colors of an A/F class main 
sequence star with absolute magnitude $M_{V0} \approx -2$. 

\subsection{X-ray pulsations}
\begin{figure}
\begin{tabular}{c}
\psfig{figure=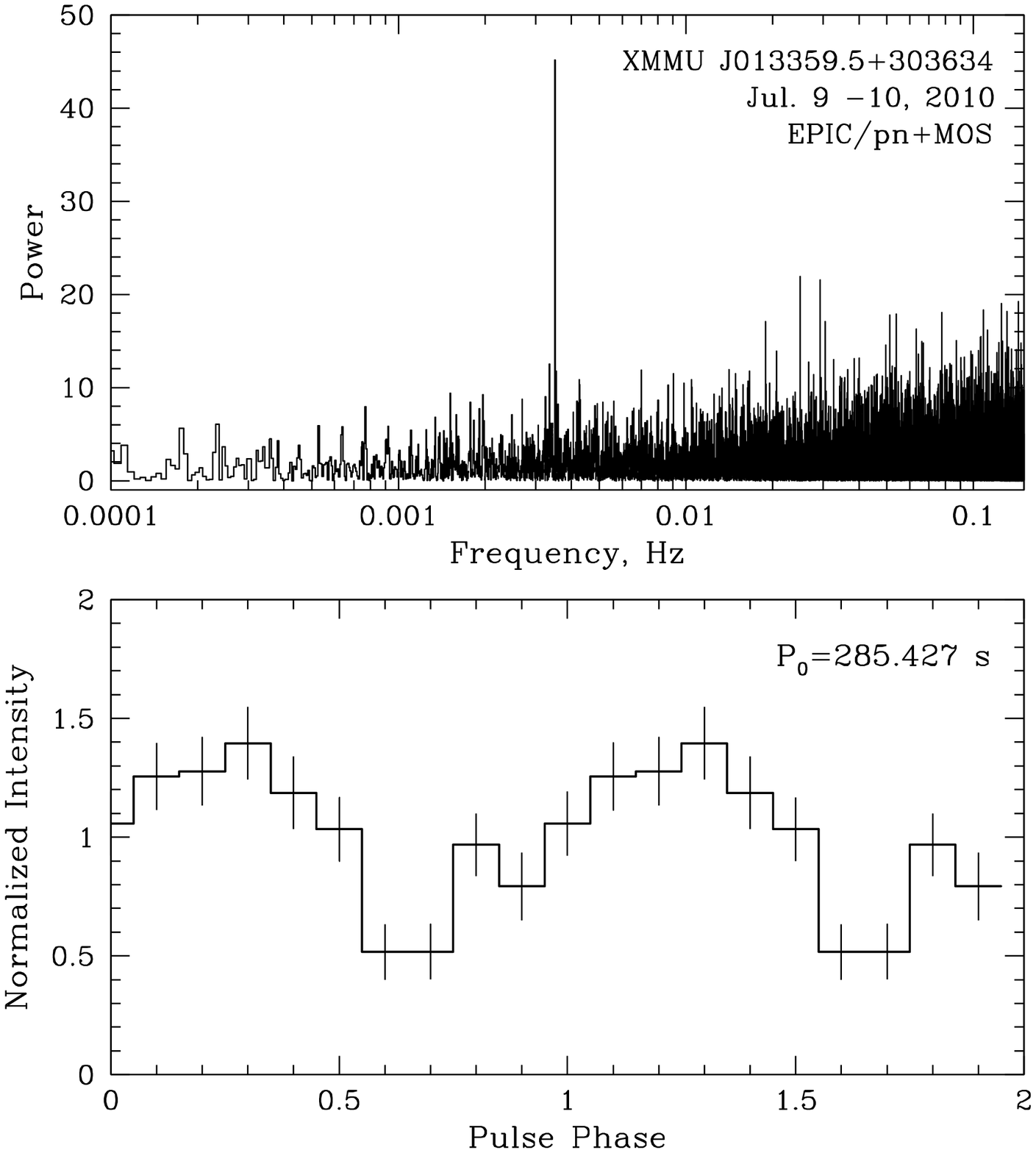,width=9.0cm,angle=0.}
\\
\psfig{figure=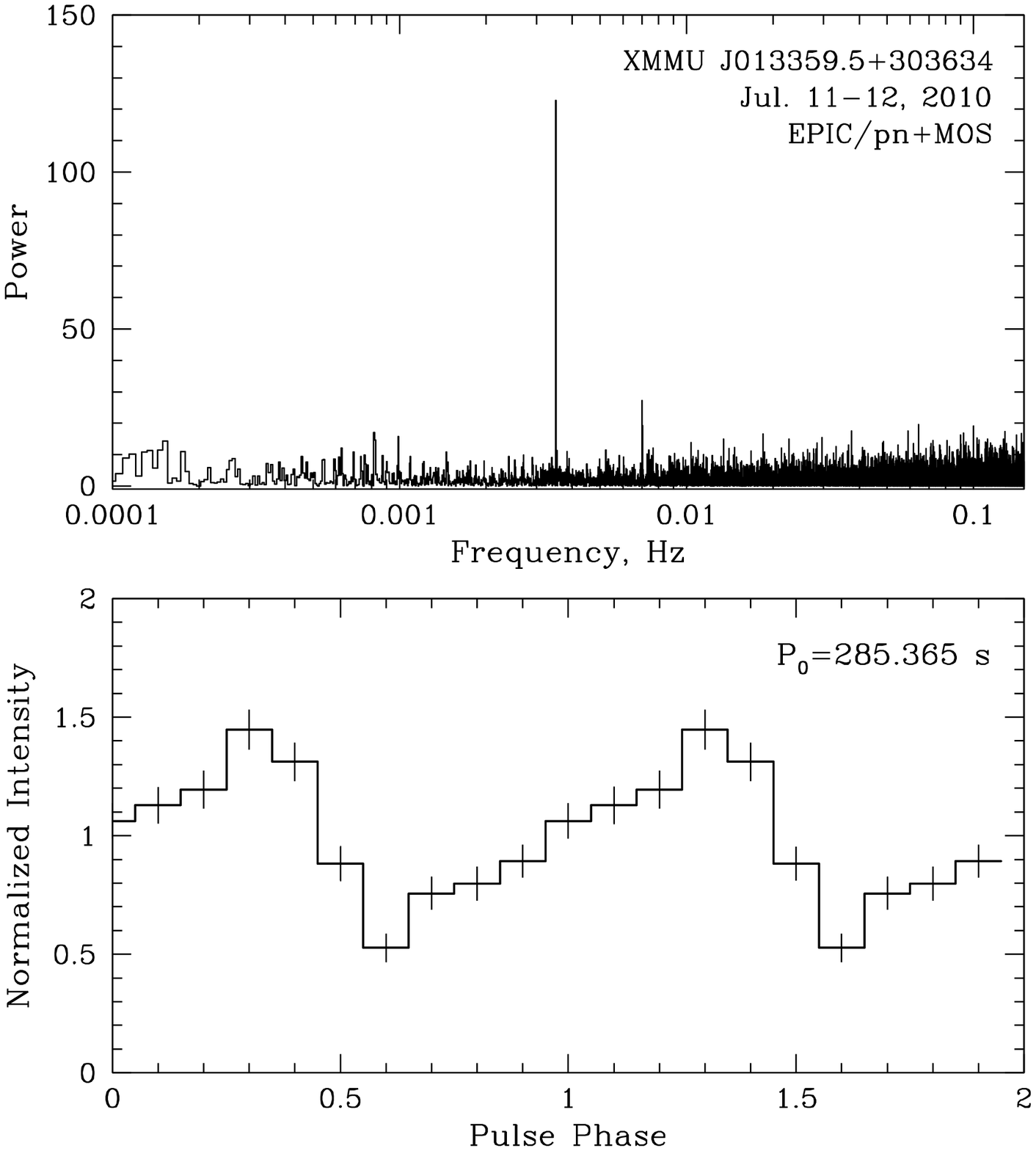,width=9.0cm,angle=0.}
\end{tabular}
\caption{Power spectra of XMMU J013359.5+303634 obtained using the data of the 2010 Jul. 9-10 and 11-12 {\em XMM-Newton}/EPIC 
(EPIC-pn, MOS1 and MOS2 detectors combined) observations in the 0.3-10 keV energy band and corresponding background corrected 
X-ray pulse profiles folded with most likely pulsation periods (Table \ref{timing_spec_par}).}
\label{pds_efold}
\end{figure}

The data from EPIC detectors in the 0.3-10 keV energy band was used to perform timing analysis of XMMU J013359.5+303634. After a barycentric 
correction of the event arrival times, I performed a Fast Fourier Transform (FFT) and Lomb-Scargle periodogram analysis, in order to search 
for coherent periodicities. The combined EPIC-pn and MOS light curves with 2.6 s time bins were used to improve sensitivity. The Fourier 
spectra and periodograms of data from the 2010 Jul. 9-10 and Jul. 11-12 {\em XMM} observations show strong peaks at the frequency of 
$\sim 3.504\times10^{-3}$ Hz (Fig. \ref{pds_efold}). The strengths of the peaks in the individual Fourier spectra (Fig. \ref{pds_efold}) 
correspond to the spurious detection probabilities of $\sim 7\times 10^{-6}$ and $\sim 3\times 10^{-23}$ for the Jul. 9-10 and Jul. 11-12 
{\em XMM-Newton} observations \citep{Vaughan94}. The source photon counts during the 2010 Aug. 12-13 observation was too low to obtain a 
significant signal in the frequency domain.

To refine the estimate of the pulsation period, a light curve folding technique was used, assuming no period change during individual 
observations. The most likely values of the period (Table \ref{timing_spec_par}) were obtained fitting the peaks in the $\chi^{2}$ versus 
trial period distribution with a combination of Gaussian function and a constant. The period errors shown in Table \ref{timing_spec_par} 
were computed following the procedure described in \citep{Leahy87}. Then, the source light curves were folded using the periods determined 
from the folding analysis. The resulting folded light curves of XMMU J013359.5+303634 in the 0.3-10 keV energy band during the first two 
observations are shown in Fig. \ref{pds_efold}. The source demonstrates quasi-sinusoidal pulse profiles with an additional significant peak 
around phase 0.3 (Fig. \ref{pds_efold}). To account for deviations from the sinusoidal shape, the harmonic representation of the folded pulse 
profiles was used (i.e. sum of harmonics plus a constant). Only the first four harmonics were kept because higher order harmonics were not 
significant. To characterize the amplitude of the pulsations, the source pulsed fraction is defined as ($I_{\rm max}-I_{\rm min}$)/($I_{\rm max}+I_{\rm min}$), 
where $I_{\rm max}$ and $I_{\rm min}$ represent the harmonic fit at the maximum and minimum of the pulse profile excluding background photons. 
The pulsed fraction was found to be stable over a period covered by 2010 Jul. 9-10 and 11-12 observations with respective amplitudes of 
46$\pm$5 and 47$\pm$3 per cent (Table \ref{timing_spec_par}). 

In order to search for a possible change in the pulsation period on the time scale of the individual observation, each observation was 
divided into a shorter 10 ks segments and timing analysis was performed on each light curve segment. In the course of both the 2010 
Jul. 9-10 and Jul. 11-12 observations, the pulsation period was stable within measurement errors. In addition, no significant change of the 
average pulsation period was detected between the 2010 Jul. 9-10 and Jul. 11-12 {\em XMM-Newton} observations (Table \ref{timing_spec_par}). 

\begin{figure}
\psfig{figure=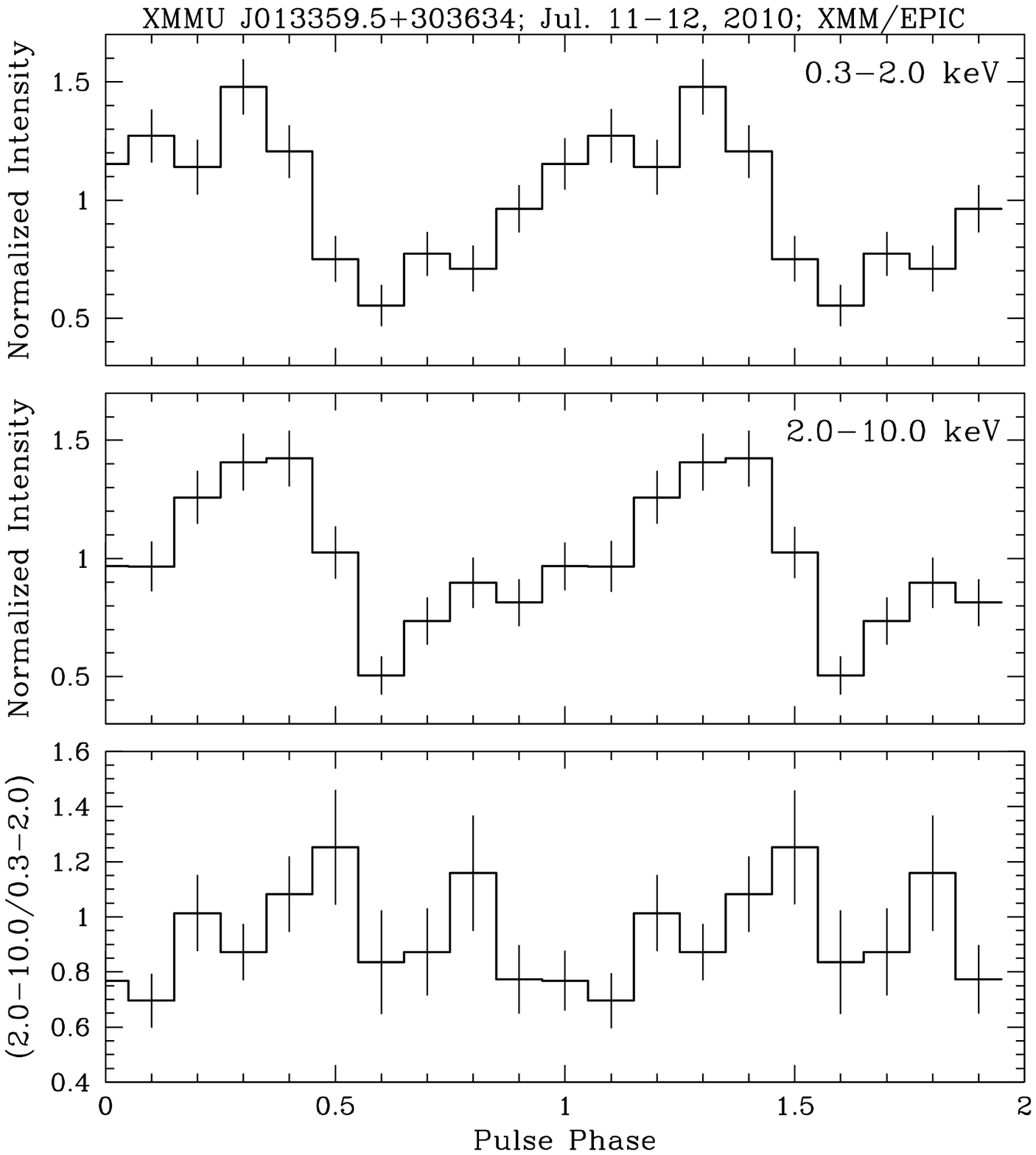,width=9.0cm,angle=0.}
\caption{Background-corrected normalized X-ray light curves of XMMU J013359.5+303634 during the 2010 Jul. 11-12 observation 
folded at the corresponding best period (Table \ref{timing_spec_par}) in the 0.3-2 and 2-10 keV energy bands 
({\em upper and middle panels}) along with their hardness ratio ({\em bottom panel}).}
\label{mod_energy_depend}
\end{figure}

No significant energy dependence of the source pulse profile was detected during the 2010 Jul. 9-12 observations. For the 2010 Jul. 11-12 
observation, pulse profiles in the soft (0.3-2 keV) and hard (2-10 keV) bands folded at the best pulsation period along with their 
ratio are shown in Fig. \ref{mod_energy_depend}. The corresponding background-corrected pulsed fractions were 46$\pm$4 for the 0.3-2 keV 
energy band and 47$\pm$4 per cent for the 2-10 keV energy band.
 
\subsection{X-ray spectra}
The pulse phase averaged {\em XMM-Newton}/EPIC spectra of XMMU J013359.5+303634 during the 2010 July 9-12 observations can be adequately 
fit with an absorbed simple power law model with photon index, $\Gamma \sim 1.2$ and an equivalent hydrogen column density 
N$_{\rm H}\sim1.5\times10^{21}$ cm$^{-2}$. The corresponding absorption-corrected luminosity of the source in the 0.3-10 keV band is 
$\sim 1.45\times10^{37}$ ergs s$^{-1}$. The best-fit spectral model parameters of the source are given 
in Table \ref{timing_spec_par}. The shape of the energy spectra of XMMU J013359.5+303634 in the 0.3-10 keV energy band did not change 
significantly between the 2010 July 9-10 and July 11-12 observations. The measured absorbing column $N_{\rm H}$ is $\sim$2.5 times higher 
than the Galactic hydrogen column in the direction of M33, 6$\times10^{20}$ cm$^{-2}$ \citep{DL90}. The excess absorbing column is 
consistent with an additional intrinsic absorption within the system and inside the disk of M33. For the 2010 Aug. 12-13 observation, 
the number of source counts was not sufficient to perform spectral analysis.

\begin{table*}
\caption{X-ray pulsation parameters and spectral fit information for XMMU J013359.5+303634. 
\label{timing_spec_par}}
\begin{tabular}{cccccccccc}
\hline
\multicolumn{3}{c}{Timing Parameters}&\multicolumn{7}{c}{POWERLAW*WABS Spectral Model Parameters}\\
\hline
Date     & Period & PF$_{0.3-10 keV}$ & N$_{\rm H}$                & Photon & Flux$^{b}$ & Flux$^{c}$ & $L_{\rm X}$$^{d}$&$L_{\rm X}$$^{e}$ & $\chi^{2}$ \\
         &  (s)   &(per cent)$^{a}$ &($\times 10^{20}$ cm$^{-2}$)& Index  &            &            &                  &                  & (d.o.f.)\\       
\hline
2010 Jul. 9-10 & $285.427\pm0.089$ & $46\pm5$ & $14\pm5$ & $1.22^{+0.13}_{-0.12}$ & $1.60\pm0.07$ & $1.76\pm0.14$      & 1.28 & 1.41 & 67.6(69)   \\
2010 Jul. 11-12& $285.365\pm0.047$ & $47\pm3$ & $15\pm2$ & $1.17\pm0.06$      & $1.69\pm0.04$ & $1.85^{+0.09}_{-0.08}$ & 1.35 & 1.48 & 190.2(143) \\
\hline
\end{tabular}

$^{a}$ -- pulsed fraction in the $0.3-10$ keV energy band\\
$^{b}$ -- absorbed model flux in the $0.3 - 10$ keV energy range in units of $10^{-13}$ erg s$^{-1}$ cm$^{-2}$\\
$^{c}$ -- unabsorbed model flux in the $0.3 - 10$ keV energy range in units of $10^{-13}$ erg s$^{-1}$ cm$^{-2}$\\
$^{d}$ -- absorbed luminosity in the $0.3 - 10$ keV energy range in units of $10^{37}$ erg s$^{-1}$, assuming the distance of 817 kpc\\
$^{e}$ -- unabsorbed luminosity in the $0.3 - 10$ keV energy range in units of $10^{37}$ erg s$^{-1}$, assuming the distance of 817 kpc\\

\end{table*}

\begin{figure}
\psfig{figure=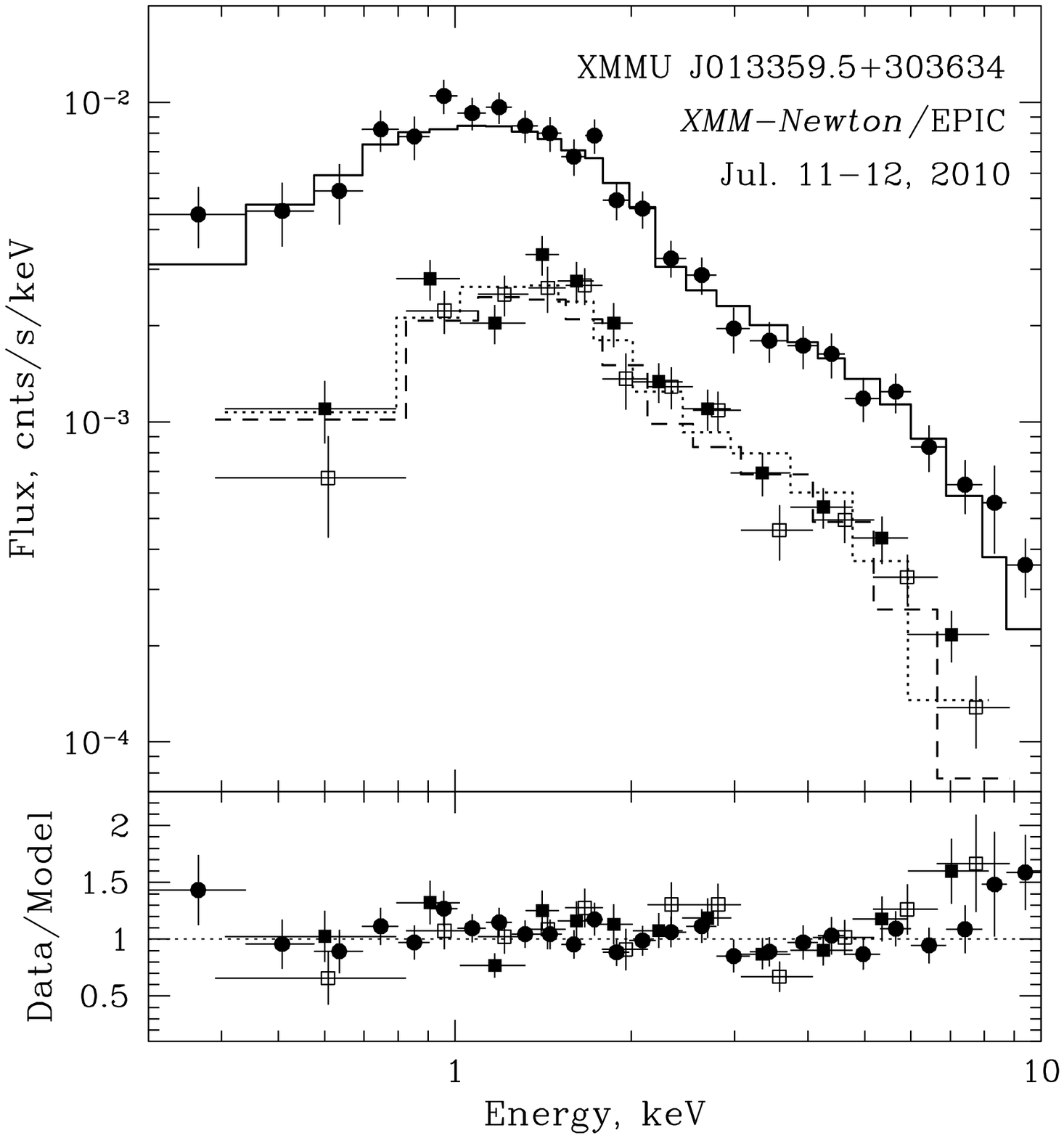,width=9.0cm,angle=0.}
\caption{EPIC count spectra and model ratios of XMMU J013359.5+303634 during the 2010 Jul. 11-12 observation. The EPIC-pn data is plotted 
with filled circles, while EPIC-MOS1 and MOS2 data is shown with filled and open squares respectively. The best-fit absorbed power law 
model approximation of EPIC-pn, MOS1 and MOS2 data is shown with solid, dotted and dashed histograms.}
\label{spec_fig}
\end{figure}

\section{Discussion and Conclusions}
The absence of a bright optical counterpart to XMMU J013359.5+303634, its overall X-ray properties and positional coincidence with one 
of the major spiral arms in M 33 (Fig. \ref{image_general}) imply that it is not a Galactic foreground object and almost certainly belongs 
to M33. The X-ray pulsations and spectra of the source are consistent with an accreting highly magnetized neutron star in a binary system.

In the high-mass system scenario, the 285.4 s pulse period places XMMU J013359.5+303634 squarely among Be systems or underfilled Roche 
Lobe supergiant systems on the Corbet diagram \citep{Corbet86,CC06}. The transient behavior of the source strongly favors the Be X-ray 
binary interpretation, in which the X-ray outbursts are attributed to either periastron passages of the compact companion (Type I outbursts) 
or changes in the Be star circumstellar disk that might occur at any binary phase (Type II outbursts) \citep{Stella86,Negueruela01}. 
Possible optical counterparts to XMMU J013359.5+303634 are all fainter than 20 mag in the V band (Table \ref{LGS_counterparts}), with absolute 
$V$ magnitudes between $\sim -2$ and $\sim -4$ at 817 kpc, supporting the Be system identification, but not ruling out a supergiant system 
with a luminosity class II companion. In addition, the brightest optical object located within the error circle of XMMU J013359.5+303634 has 
properties consistent with being an early B V star and possibly belonging to the Be class. Assuming the distance of 817 kpc, the 
absorption-corrected $0.3-10$ keV X-ray luminosity of XMMU J013359.5+303634 measured in the Jul. 9-12 observations 
($L_{X}\sim 1.45 \times 10^{37}$ ergs s$^{-1}$) also falls within the range observed for Be X-ray binary pulsars and exceeds typical 
luminosities observed in the wind-fed supergiant systems \citep{Bildsten97,CC06}. 

If XMMU J013359.5+303634 is indeed a Be system, its binary period, $P_{orb}$ estimated from the Corbet diagram falls in the $\sim 100-200$ 
day range. The observed time span between {\em XMM-Newton} observations when the source was detected allows to put an additional limit 
on the orbital period. Assuming that a single Type I outburst was observed and that it was lasting less than one orbital cycle requires 
$P_{orb} > 35$ days.   

Another possibility is that XMMU J013359.5+303634 is a transient low-mass binary system \citep{Bildsten97}. In that scenario, the transient 
outbursts could be explained as a result of a viscous-thermal instability of the quiescent accretion disk. The relatively long pulsation 
period of the source almost certainly excludes low-mass short orbital period systems. On the other hand, the possibility of a very long 
period system with red giant companion similar to the Galactic source GX 1+4 \citep{CR97} can not be completely ruled out, since the expected 
brightness of the red giant companion ($M_{V}\gtrsim -1$ mag) does not exceed the brightness of the individual optical objects detected within 
the error circle of XMMU J013359.5+303634. 

Although most of the observed properties of XMMU J013359.5+303634 favor a Be/X-ray binary identification, a proper optical identification is 
essential to determine the actual nature of the system. To facilitate it, deeper high-resolution optical observations of the source region 
are needed along with an improved X-ray position measurement. Regular monitoring X-ray observations of the central region of M33 have a 
potential to detect new outbursts from XMMU J013359.5+303634, improve source localization and put constraints on its orbital parameters.      

The number of X-ray pulsar candidates with non-supersoft spectra located outside our own Galaxy and Magellanic Clouds is still very small 
with only four such systems known so far, including XMMU J013359.5+303634. All of them were discovered as transient sources with X-ray 
luminosities in the $L_{X}\sim 10^{37} - 10^{39}$ ergs s$^{-1}$ range, and probably belong to a luminous Be/X-ray binary class. Because of 
their higher mass transfer rates, luminous X-ray pulsars are ideal laboratories to study effects of angular momentum transfer and interaction 
between pulsar magnetosphere and the accretion flow through the evolution of their pulsation periods \citep{RJ77,GL79}. Analysis of the pulse 
profiles and spectral properties of these objects can provide extremely valuable information on the accretion flow and magnetic field geometry, 
neutron star compactness and emission properties \citep{BS76,AP10}. The discovery of XMMU J013359.5+303634 is yet another demonstration that 
monitoring observations with current X-ray missions can detect these and similar pulsating sources (both transient and persistent) in the 
galaxies located up to a few Mpc away. Observations of nearby galaxies allow us to monitor a significantly bigger X-ray source population 
when compared to just our own Galaxy and its immediate neighbors. In the galaxies with recent and ongoing star formation, X-ray pulsars 
should constitute a sizable fraction of the total X-ray source population, with some of them expected to be bright enough ($L_{X}\gtrsim 10^{37}$ 
ergs s$^{-1}$) for detection by current instruments, opening a way to significantly increase the statistics of luminous X-ray pulsars.  

\section*{Acknowledgments}
The author would like to thank the referee for comments and suggestions that helped to improve the paper. This research has made use of data 
obtained through the High Energy Astrophysics Science Archive Research Center Online Service, provided by the NASA/Goddard Space Flight Center. 
XMM-Newton is an ESA Science Mission with instruments and contributions directly funded by ESA Member states and the USA (NASA). {\em Chandra} 
X-ray observatory is operated by the Smithsonian Astrophysical Observatory on behalf of NASA. This research also made use of NASA/IPAC 
Extragalactic Database (NED), which is operated by the Jet Propulsion Laboratory, California Institute of Technology, under contract with NASA.

\end{document}